\documentclass{PoS}
 
\usepackage{subfigure}
\usepackage{amsmath}

\title{Studies of electroweak boson production in the forward region with LHCb}

\ShortTitle{Studies of electroweak boson production in the forward region with LHCb}

\author{\speaker{James Keaveney (on behalf of the LHCb collaboration}\thanks{The author would like to acknowledge the support of Science Foundation Ireland}%
        \\
       University college Dublin\\
       E-mail: \email{james.keaveney@ucdconnect.ie}}

\abstract{Studies of electroweak boson production at LHCb are detailed and discussed.  Proposed signal selection
schemes and background suppression strategies are described and the projected performance is estimated using
Monte Carlo data. Due to the unique pseudorapidity coverage and triggering capabilities of LHCb, 
these studies will probe an unexplored region of  $(x, Q^2)$ space.}

\FullConference{XVIII International Workshop on Deep-Inelastic Scattering and Related Subjects\\
		 April 19 -23, 2010\\
		 Convitto della Calza, Firenze, Italy}

\begin{document}

\section{Introduction}
The LHCb experiment \cite{alves}, one of the four main experiments at the LHC, has begun taking data at $\sqrt{s}$ =  7TeV and expects to take 1fb$^{-1}$ by the end of 2011.
Studies of electroweak boson production in this data form a significant component of the early physics programme at the LHC for numerous reasons.
Firstly, these studies can make a precise test of the standard model  at a new energy regime and provide an input to proton PDF constraining procedures. 
Secondly, the low theoretical uncertainty on the Z boson production cross section coupled with the ability to select signal samples of high purity makes an indirect luminosity measurement possible with a relatively small data sample.  Thirdly, the clean experimental signature of the muonic decay channels allows detector calibration and 
performance studies to be performed using early data.

As electroweak theory can currently describe the fundamental partonic processes of electroweak boson production at the LHC at next-to-next-to-leading order \cite{melnikov}, 
	the dominant uncertainties on theoretical predictions of these processes 
  arise from the knowledge of proton PDFs. In kinematic regions where PDF uncertainties are low,  precise measurements of electroweak bosons 
  provide a stringent test of the standard model. In other regions where PDF uncertainties are large, these studies can constrain PDFs.  Uniquely 
   among the LHC experiments, LHCb is instrumented in one direction only, with acceptance in the pseudorapidity region $1.9 < \eta< 4.9$. 
   This angular acceptance is  complimented by the ability to trigger and reconstruct muons in the kinematic range ($P_{\mu} > 6$ GeV, $Pt_{\mu} > 1$ GeV).
   By studying the low mass $\gamma*\rightarrow -> \mu^+\mu^-$  LHCb can access  two distinct regions of $(x, Q^2)$ space, one of which is previously unexplored. 

We present the methods developed to select  samples of $W\rightarrow\mu{\nu},  Z\rightarrow\mu^+\mu^- $ and $\gamma* \rightarrow \mu^+\mu^-$ events. The
   efficiencies, purities and yields of these methods are estimated using Monte Carlo data. 
   
\section{Triggering and offline selection of signal events}
\subsection{$W\rightarrow\mu\nu$ selection}
 The muonic decay channel of the W boson is characterized by a single, isolated muon with a large transverse momentum. At LHCb, these events will be initially
 triggered by requiring a muon  with $P_{T} > 1$ GeV in the Level-0 hardware trigger. In the subsequent software stages of the LHCb trigger we 
 require a muon with $P_{T} > 20$ GeV to be reconstructed. The largest backgrounds to these events arise from the semi-leptonic decay of heavy quark
  hadrons, $Z\rightarrow\mu^+\mu^- $ events in which one muon decays outside the LHCb acceptance and mis-identification of high momentum hadrons as muons.
  To suppress these backgrounds offline, the candidate muon track is required to have  $P_{T} > 30$ GeV. Isolation requirements are then imposed on the remaining
  candidate tracks. 
In Equation \ref{eq:solve} we define an asymmetry, $A_{pt}$, 
between the $P_{T}$ of the candidate track and the summed $P_{T}$ of all other tracks
in the event and an asymmetry $C_{pt}$ between the $P_{T}$ of the candidate track and the summed $P_{T}$ of all tracks
in an imaginary cone around the candidate of radius 1 in eta-phi space,
\begin{align}\label{eq:solve}
  A_{pt} =  \frac{Pt_{\mu}  - Pt_{other \: tracks \:in \;event} }{ Pt_{\mu}  + Pt_{other\; tracks\; in\; event} }
  \hspace{1.5cm}
  C_{pt} =  \frac{Pt_{\mu}  - Pt_{all\; tracks \;in \;cone} }{ Pt_{\mu}  + Pt_{all\; tracks \;in \;cone} }
\end{align}
and require  $A_{pt}>0.3 and C_{pt}>0.7$ 

\begin{figure}[ht]
\centering
\subfigure[Before selection cuts]{
\includegraphics[scale=.342]{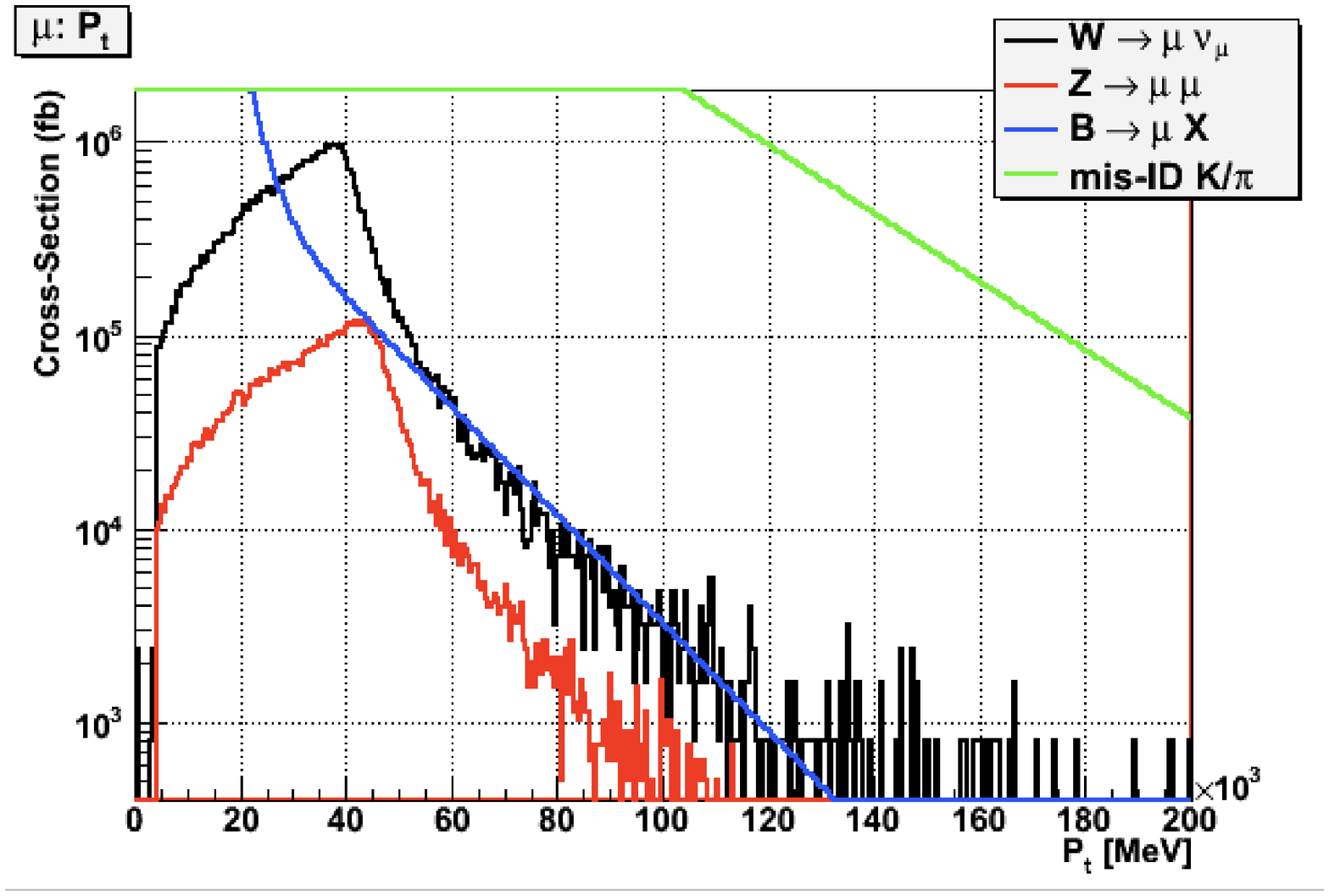}
\label{fig:Wselscheme}
}
\subfigure[After selection cuts]{
\includegraphics[scale=.454]{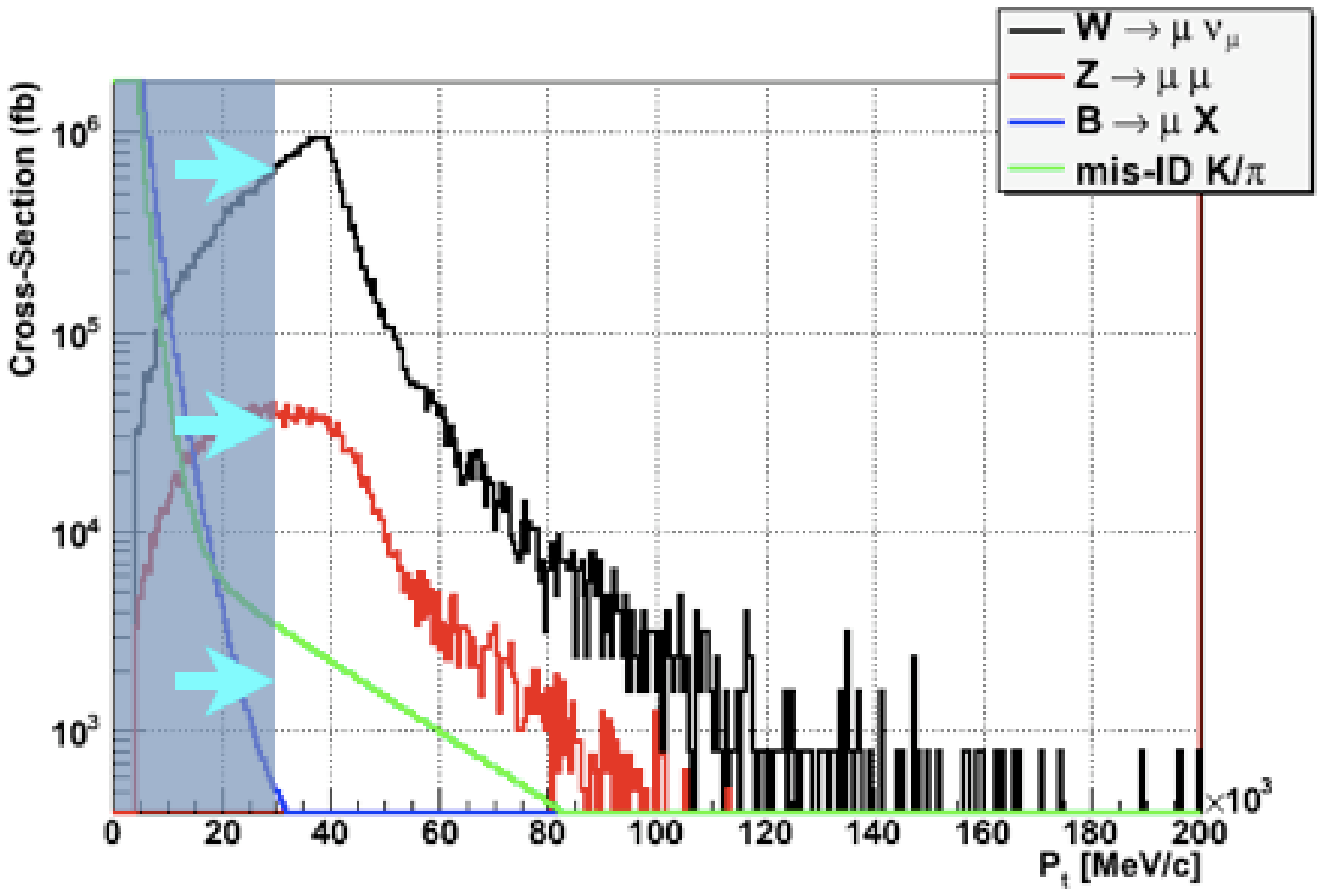}
\label{fig:subfig2}
}
\label{fig:subfigureExample}
\caption[Optional caption for list of figures]{$Pt_{\mu} $  distributions for  $W\rightarrow\mu{\nu}$ and background components before and after selection criteria are applied. The $Pt_{\mu}  > 30$ GeV requirement is illustrated.  }
\end{figure}

In Fig.~\ref{fig:subfigureExample} the effect of the selection cuts on signal and background is shown. 


\subsection{Z and $\gamma*$   selection}
Similar techniques  are used to the select dimuons arising from the decays of Z and $\gamma*$ bosons. 
These events are triggered by the requirement of a di-muon with  $P_{T} > 1.5$ GeV  in the initial L0 hardware trigger and requirement 
of a dimuon with an invariant mass in the range  $2.5< m_{\mu\mu} < 40$ GeV  for  $\gamma*\rightarrow\mu^+\mu^- $ events  and  $m_{\mu\mu} > 40$ GeV for $Z\rightarrow\mu^+\mu^- $ events in the subsequent software trigger.
 Backgrounds are dominated by the decays of heavy quarks and the mis-ID of high momentum di-hadrons as di-muons. In the offline selection of Z bosons, requiring  two
muons that originated at the primary vertex, each with  $P_{T} > 20$ GeV and the hadronic energy associated to each track <50GeV with  
$71< m_{\mu\mu} < 111$ GeV, yields a selected signal sample with 97\% purity and 91\% efficiency on triggered events \cite{anderson}. To further suppress backgrounds 
and increase selected signal sample purity the isolated nature of muon tracks from electroweak boson decay is exploited.
For a given dimuon candidate, the cone isolation ($C_{pt} $, previously defined in (\ref{eq:solve})) of each muon is used to construct a dimuon asymmetry $I_{pt}$
\begin{equation}\label{eq:Di}
  I_{pt} = \sqrt{ \frac{ ( C_{pt}^{\mu+} -1)^{2}   +   ( C_{pt}^{\mu-} -1)^{2} }{ 8} }
\end{equation}
 By requiring dimuon candidates to have  $I_{pt}  < 0.2 $, the purity of the selected sample is increased to 99\%. 
 The background components described are far more dominant at the low invariant mass region corresponding to $ \gamma*\rightarrow\mu^+\mu^- $
events. To suppress this large background the candidate tracks are required to pass a pre-selection based on impact parameter significance
(IPS < 3) and track $\chi^{2}$ per degree of freedom less than 2.5. A muon hypothesis likelihood is constructed from subdetector
information associated to a muon candidate and a requirement is placed on this variable to reject background arising from mis-identified pions and kaons.
Finally, a Fisher discriminant is constructed from four asymmetries
of the form  $A(x,y)   =  (x-y)/(x+y)$:
\begin{align}
A(P_{\mu^{+}} , P_{cone^{+}}),  
\hspace{.4cm}
A(P_{\mu^{-}} , P_{cone^{-}}) , 
\hspace{.4cm}
A(P_{\mu^{+}} +  P_{\mu^{-}}  , P_{other\; tracks\; in\; event}),
\hspace{.4cm}
A(P_{cone^{+}} +  P_{cone^{-}}  , P_{other\; tracks\; in\; event}),
\end{align}
and cuts on the Fisher discriminant are optimized for purity in four distinct regions of the  $\gamma*\rightarrow\mu^+\mu^-$ mass spectrum \cite{andersonDIS}.

\subsection{Expected yields, efficiencies, purities and uncertainties}
The selection schemes described above have been tested using Monte Carlo data and the full LHCb detector simulation framework.
In Table~\ref{tab:PPer}, the yields , efficiencies and purities are estimated. Systematic errors have been estimated for size of the background components,
reconstruction efficiencies, trigger efficiencies and integrated lumnosity. With 1fb$^{-1}$ of data, it is forseen that uncertainties will be dominated 
by the knowledge of the integrated luminosity.

\begin{table}[h!] 
\caption{Estimated performances of the selection schemes }
\centering       

\begin{tabular}{l  rrrr}
\\ [0.2ex]  
\hline\hline                         
 Channel  &\#       events in LHCb in 1 fb$^{-1}$ &  total efficiency  & \# events recorded  &  purity  
\\ [0.25ex]    
\hline                  
 
$W\rightarrow\mu{\nu}$ &   3470000  &.81 &2810000&.94 \\[-1ex] 
\\ [0.2ex]    
 
 \hline
 
$Z\rightarrow\mu^+\mu^- $  &   170000  & .79 &               134300 & .99\\[-1ex] 
\\ [0.2ex]   
\hline

$ \gamma*\rightarrow\mu^+\mu^- $ &   124000  & .19   & 23732 & .95\\[-1ex] 
($2.5<m_{\mu\mu}<5$GeV)
\\ [0.2ex]

\hline

$ \gamma*\rightarrow\mu^+\mu^-$ &   154835  &  .37 & 57289 &.95\\[-1ex] 
($5<m_{\mu\mu}<10$GeV)
\\ [0.2ex]

\hline
$ \gamma*\rightarrow\mu^+\mu^- $ &   75023  &.39  &29259&.95\\[-1ex] 
$(10<m_{\mu\mu}<20$ GeV)
\\ [0.2ex]  

\hline
$ \gamma*\rightarrow\mu^+\mu^- $ &   21549  & .39 &8404&.95\\[-1ex] 
     ( $20<m_{\mu\mu}<40$ GeV)
   \\ [0.25ex]

\hline                          
\end{tabular} 
\label{tab:PPer} 
\end{table}

\section{Conclusions}
In 2010 and 2011, the LHCb experiment will collect  ~1fb$^{-1}$ of data. Analysis techniques to select the muonic decays
of electroweak bosons in this dataset with high purity have been developed. These studies have numerous motivations: testing of the standard model; 
 constrainment of proton PDFs in an unexplored area of ($x, Q^2$) space; integrated luminosity estimation and detector calibration.

\end{document}